\documentclass[runningheads]{llncs}
\usepackage{lineno}
\usepackage{cite}
\usepackage{amsmath,amssymb,amsfonts}
\usepackage{algorithmic}
\usepackage{graphicx}
\usepackage{textcomp}
\usepackage{xcolor}
\usepackage{tikz}
\usepackage{listings,multicol}
\usepackage[skins]{tcolorbox}
\usepackage[export]{adjustbox}
\usepackage{ragged2e}
\usepackage{fancyvrb}
\usepackage{verbatimbox}
\usepackage{etoolbox}
\usepackage{float, xspace, hyperref, makecell,tabularx, xurl}
\usepackage{svg}
\usetikzlibrary{shapes,arrows,positioning,calc}

\newcommand{\aesprf}{\textsc{aes}\_\textsc{cmac}\_\textsc{prf}\xspace}
\newcommand{\meshcopfull}{Mesh Commissioning Protocol\xspace}

\newcommand{\zkp}{Zero knowledge proof\xspace}
\newcommand{\dtls}{\textsc{dtls}\xspace}

\newcommand{\bas}{\textsc{bas}\xspace}

\begin{document}
	%
	
	\title{Symbolic Security Verification of \meshcopfull in Thread (extended version)}
	%
	%
	\author{Pankaj Upadhyay \inst{1,2,3}  
		Subodh Sharma\inst{2} \and
		Guangdong Bai\inst{3}
	}
	%
	%
	\institute{The University of Queensland and Indian Institute of Technology Delhi Academy of Research \and
		Indian Institute of Technology Delhi, India\\
		\email{\{qiz218586,svs\}@iitd.ac.in}\\
		\and
		The University of Queensland, Australia\\
	\email{g.bai@uq.edu.au}}
\maketitle              

\newcommand{\commpetreq}{\textsc{comm}\textunderscore \textsc{pet}.req\xspace}
\newcommand{\commpetres}{\textsc{comm}\textunderscore \textsc{pet}.rsp\xspace}
\newcommand{\commkareq}{\textsc{comm}\textunderscore \textsc{ka}.req\xspace}
\newcommand{\commkares}{\textsc{comm}\textunderscore \textsc{ka}.rsp\xspace}
\newcommand{\leadpetreq}{\textsc{lead}\textunderscore \textsc{pet}.req\xspace}
\newcommand{\leadpetres}{\textsc{lead}\textunderscore \textsc{pet}.rsp\xspace}
\newcommand{\leadkareq}{\textsc{lead}\textunderscore \textsc{ka}.req\xspace}
\newcommand{\leadkares}{\textsc{lead}\textunderscore \textsc{ka}.rsp\xspace}
\newcommand{\mgmtgetreq}{\textsc{mgmt}\textunderscore \textsc{get}.req\xspace}
\newcommand{\mgmtgetres}{\textsc{mgmt}\textunderscore \textsc{get}.rsp\xspace}
\newcommand{\mgmtsetreq}{\textsc{mgmt}\textunderscore \textsc{set}.req\xspace}
\newcommand{\mgmtsetres}{\textsc{mgmt}\textunderscore \textsc{set}.rsp\xspace}
\newcommand{\joinfinreq}{\textsc{join}\_\textsc{fin}.req\xspace}
\newcommand{\joinfinres}{\textsc{join}\textunderscore \textsc{fin}.rsp\xspace}
\newcommand{\disreq}{\textsc{dis\textunderscore req \xspace}}
\newcommand{\disres}{\textsc{dis\textunderscore res \xspace}}
\newcommand{\dtlsccf}{\textsc{dtls\textunderscore cke\textunderscore ccs\textunderscore fin}\xspace}
\newcommand{\dtlscf}{\textsc{dtls\textunderscore ccs\textunderscore fin}\xspace}
\newcommand{\dtlsshskeshd}{\textsc{dtls\textunderscore sh\textunderscore ske\textunderscore shd}\xspace}
\newcommand{\dtlsch}{\dtls ClientHello\xspace}
\newcommand{\dtlshvr}{\dtls HelloVerifyReq\xspace}
\newcommand{\dtlschwc}{\dtls ClientHello WC\xspace}
\newcommand{\dtlsske}{\dtls ServerKeyExchange\xspace}
\newcommand{\dtlsshd}{\dtls ServerHelloDone\xspace}
\newcommand{\dtlscke}{\dtls ClientKeyExchange\xspace}
\newcommand{\dtlsccs}{\dtls ChangeCipherSpec\xspace}
\newcommand{\dtlsfin}{\dtls Finished\xspace}
\newcommand{\kektlv}{\textsc{kektlv}\xspace}

\newcommand{\typeccs}{change\textunderscore cipher\textunderscore spec\xspace}
\newcommand{\typedtlsfin}{dtls\textunderscore finished\xspace}
\newcommand{\typeleadkareq}{\textsc{leadkareq}\xspace}
\newcommand{\typemgmtrq}{\textsc{mgmtrq}\xspace}
\newcommand{\pskc}{\textsc{psk}c\xspace}
\newcommand{\ba}{\textsc{ba}\xspace}
\newcommand{\cm}{\textsc{cm}\xspace}
\newcommand{\ld}{\textsc{ld}\xspace}
\newcommand{\jo}{\textsc{jo}\xspace}
\newcommand{\jr}{\textsc{jr}\xspace}
\newcommand{\br}{\textsc{br}\xspace}

\newcommand{\funcdtlsfin}{senc\textunderscore dtls\textunderscore finished\xspace}
\newcommand{\funcgenkek}{\textsc{genkek}\xspace}
\newcommand{\funczkp}{\textsc{zkp}\xspace}
\newcommand{\names}{\emph{Names}\xspace}

\newcommand{\param}[1]{\emph{#1}}
\newcommand{\process}[1]{\textsc{#1}}
\newcommand{\calA}{\mathcal{A}}
\newcommand{\calB}{\mathcal{B}}
\newcommand{\calC}{\mathcal{C}}
\newcommand{\calD}{\mathcal{D}}
\newcommand{\calE}{\mathcal{E}}
\newcommand{\calF}{\mathcal{F}}
\newcommand{\calG}{\mathcal{G}}
\newcommand{\calH}{\mathcal{H}}
\newcommand{\calK}{\mathcal{K}}
\newcommand{\calM}{\mathcal{M}}
\newcommand{\calP}{\mathcal{P}}
\newcommand{\calR}{\mathcal{R}}
\newcommand{\calS}{\mathcal{S}}
\newcommand{\calT}{\mathcal{T}}
\newcommand{\calX}{\mathcal{X}}
\newcommand{\calY}{\mathcal{Y}}

\newcommand{\qcookiesa}{\textsc{q\textunderscore cookies\textunderscore 1}\xspace}
\newcommand{\qcookiesb}{\textsc{q\textunderscore cookies\textunderscore 2}\xspace}
\newcommand{\qkeyderiv}{\textsc{q\textunderscore pskc\textunderscore deriv}\xspace}
\newcommand{\qcredtrans}{\textsc{q\textunderscore cred\textunderscore trans}\xspace}
\newcommand{\qcredtransauth}{\textsc{q\textunderscore cred\textunderscore transseq}\xspace}
\newcommand{\qsamessk}{\textsc{q\textunderscore same\textunderscore ssk}\xspace}
\newcommand{\qsamesskcm}{\textsc{q\textunderscore same\textunderscore ssk\textunderscore comm}\xspace}
\newcommand{\qsamedskj}{\textsc{q\textunderscore same\textunderscore dskj}\xspace}
\newcommand{\qcommautha}{\textsc{q\textunderscore comm\textunderscore auth\textunderscore 1}\xspace}
\newcommand{\qcommauthb}{\textsc{q\textunderscore comm\textunderscore auth\textunderscore 2}\xspace}
\newcommand{\qjoinauthc}{\textsc{q\textunderscore join\textunderscore auth\textunderscore 3}\xspace}
\newcommand{\qjoinauthd}{\textsc{q\textunderscore join\textunderscore auth\textunderscore 4}\xspace}
\newcommand{\qcommpet}{\textsc{q\textunderscore comm\textunderscore pet}\xspace}
\newcommand{\qcommka}{\textsc{q\textunderscore comm\textunderscore ka}\xspace}
\newcommand{\qleadpet}{\textsc{q\textunderscore lead\textunderscore pet}\xspace}
\newcommand{\qleadka}{\textsc{q\textunderscore lead\textunderscore ka}\xspace}
\newcommand{\qreachability}{\textsc{q\textunderscore reachability}}

\newcommand{\eventrcvdckfrmba}{cmc\textunderscore rcvd\textunderscore ck\textunderscore frm\textunderscore ba\xspace}
\newcommand{\eventsntcktocmc}{snt\textunderscore ck\textunderscore to\textunderscore cmc\xspace}
\newcommand{\eventbgnxxx}{bgn\textunderscore xxx\xspace}
\newcommand{\eventendxxx}{end\textunderscore xxx\xspace}
\newcommand{\eventjoclrcvdckfrmcms}{jo\textunderscore cl\textunderscore rcvd\textunderscore ck\textunderscore frm\textunderscore cms\xspace}
\newcommand{\eventsntcktojo}{snt\textunderscore ck\textunderscore to\textunderscore jo\xspace}
\newcommand{\eventldrcvptrqfrmba}{ld\textunderscore rcvptrq\textunderscore frm\textunderscore ba\xspace}
\newcommand{\eventauthcmba}{auth\textunderscore \cm\textunderscore \ba\xspace}
\newcommand{\eventbasntptrqtold}{ba\textunderscore sntptrq\textunderscore to\textunderscore ld\xspace}
\newcommand{\eventjrrcvnrcrds}{jr\textunderscore rcv\textunderscore nrcrds\xspace}
\newcommand{\eventjrsndnrcrds}{jrt\textunderscore snd\textunderscore nrcrds\xspace}
\newcommand{\eventcmssndkekjrt}{cmssndkekjrt\xspace}
\newcommand{\eventjtrrcvkekcms}{jtrrcvkekcms\xspace}


\newcommand{\OTHERVARS}{\textsc{othervars}\xspace}
\newcommand{\PKDF}{\textsc{pkdf2}\xspace}
\newcommand{\JOINFINRES}{\textsc{join}\textunderscore \textsc{fin}\_\textsc{res}\xspace}
\newcommand{\JOINFINREQ}{\textsc{join}\_\textsc{fin}\_\textsc{req}\xspace}
\newcommand{\LEADKARQ}{\textsc{leadkarq}\xspace}
\newcommand{\LEADKARS}{\textsc{leadkars}\xspace}
\newcommand{\LEADPETRQ}{\textsc{leadpetrq}\xspace}
\newcommand{\LEADPETRS}{\textsc{leadpetrs}\xspace}
\newcommand{\COMMSESSION}{\textsc{comm}\textunderscore \textsc{session}\xspace}
\newcommand{\COMMSESSIONRCV}{\textsc{comm}\textunderscore \textsc{sessionrcv}\xspace}

\newcommand{\eventBSrvsharedsecretkey}{bsrvssk\xspace}
\newcommand{\eventCClisharedsecretkey}{cclissk\xspace}
\newcommand{\eventbsrvfin}{bsrvfin\xspace}
\newcommand{\eventcclibeg}{cclibeg\xspace}
\newcommand{\eventcclifin}{cclifin\xspace}
\newcommand{\eventbsrvbeg}{bsrvbeg\xspace}
\newcommand{\eventrcvcommresponse}{rcvcommrsp}
\newcommand{\eventsntcommresponse}{sntcommrsp}
\newcommand{\eventleaderrep}{leaderrep}
\newcommand{\eventrcvcommkaresponse}{rcvcommkarsp\xspace}
\newcommand{\eventsntcommkaresponse}{sntcommkarsp\xspace}
\newcommand{\eventleaderrepka}{leaderrepka\xspace}

\newcommand{\eventdskjnr}{eventdskjnr\xspace}
\newcommand{\eventdskcmm}{eventdskcmm\xspace}
\newcommand{\eventjoinerrcvck}{joinerrcvck\xspace}
\newcommand{\eventcclircvck}{cclircvck\xspace}
\newcommand{\eventbsrvsntck}{bsrvsntck\xspace}
\newcommand{\eventjoinersntck}{joinersntck\xspace}
\newcommand{\eventcsrvsntck}{csrvsntck\xspace}
\newcommand{\eventjoinerfin}{joinerfin\xspace}
\newcommand{\eventcsrvbeg}{csrvbeg\xspace}
\newcommand{\eventcsrvfin}{csrvfin\xspace}
\newcommand{\eventjoinerbeg}{joinerbeg\xspace}
\newcommand{\joinergtsnetcreds}{joinergtsnetcreds}
\newcommand{\evjrtrsendsnetcreds}{evjrtrsendsnetcreds}
\newcommand{\eventCSrvsharedsecretkey}{csrvssk\xspace}
\newcommand{\eventJoinersharedsecretkey}{joinerssk\xspace}

\newcommand{\channelchbtoL}{chBtoL\xspace}
\newcommand{\channelchrtob}{chrtob\xspace}
\newcommand{\channelchbtoc}{chbtoc\xspace}
\newcommand{\channelchjtoc}{chjtoc\xspace}
\newcommand{\channelch}{ch\xspace}

\newcommand{\queryitem}[1]{{\bf{#1}}}
\newcommand{\eventname}[1]{{\bf{#1}}}

\newcommand{\CLIDH}{\textsc{clid}\_\textsc{h}\xspace}
\newcommand{\SRIDH}{\textsc{srid}\_\textsc{h}\xspace}
\newcommand{\JNIDH}{\textsc{jnid}\_\textsc{h}\xspace}
\newcommand{\SRID}{\textsc{srid}\xspace}
\newcommand{\ID}{\textsc{id}}
\newcommand{\NETCREDS}{\textsc{netcreds}\xspace}
\newcommand{\GA}{$\textsc{g}_1$\xspace}
\newcommand{\XA}{$\textsc{x}_1$\xspace}
\newcommand{\XB}{$\textsc{x}_2$\xspace}
\newcommand{\XC}{$\textsc{x}_3$\xspace}
\newcommand{\XD}{$\textsc{x}_4$\xspace}
\newcommand{\protocolcomm}{\textit{Commissioner Protocol}\xspace}
\newcommand{\protocolthreadmanagement}{\textit{Thread Management Protocol}\xspace}
\newcommand{\protocoljoiner}{\textit{Joiner Protocol}\xspace}

\newcommand{\authnote}[2]{{\bf \textcolor{blue}{#1}: \em \textcolor{red}{#2}}}
\newcommand{\baigd}[1]{\authnote{GBai}{#1}}
\newcommand{\pnkjnote}[1]{{\bf \textcolor{teal}{#1}}}

\setlength{\textfloatsep}{10.0pt plus 1.0pt minus 2.0pt}

\begin{abstract}
	The Thread protocol (or simply \emph{Thread}) is a popular networking protocol for the Internet of Things (IoT). 
	It allows seamless integration of a set of applications and protocols, hence reducing the risk of incompatibility among different applications or user protocols.
	Thread has been deployed in many popular smart home products by the majority of IoT manufacturers, such as Apple TV, Apple HomePod mini, eero 6, Nest Hub, and Nest Wifi.
	Despite a few empirical analyses on the security of Thread, there is still a lack of formal analysis on this infrastructure of the booming IoT ecosystem.
	In this work, we performed a formal symbolic analysis of the security properties of Thread.
	Our main focus is on MeshCoP (\meshcopfull), the main sub-protocol in Thread for secure authentication and commissioning of new, untrusted devices inside an existing Thread network. 	This case study presents the challenges and proposed solutions in modeling  MeshCoP.
	We use ProVerif, a symbolic verification tool of $\pi$-calculus models, for verifying the security properties of MeshCoP.
	
	\keywords{Symbolic Analysis \and Thread \and \meshcopfull.}
\end{abstract}
\section{Introduction}

In recent years, the Internet of Things (IoT) has emerged as a promising technology to enhance ease of living.
It plays significant roles in various domains, such as automobiles, agriculture, education, and smart homes.
Nearly 25 billion IoT devices are expected to exist in 2030 \cite{bgartner}.  
An IoT system typically includes multiple devices manufactured and developed by different manufacturers, and such heterogeneity often leads to incompatibility issues.
To handle this, the Thread protocol \cite{bthread} (or Thread), a wireless mesh networking protocol, is developed to provide reliable device-to-device communication.

Thread is primarily concerned with the networking layer, facilitating any IP-based application to use its functionalities later.
It uses Internet Protocol version 6 (IPv6) to handle absolute addressing and a network-wide key to protect IEEE 802.15.4 \cite{bieee} MAC data frames for security purposes.
It also supports various upper-layer protocols and IPv4 compatibility.
Thread has been deployed in various popular home automation devices, such as Google Nest, Nanoleaf Smart devices, evehome appliances, Samsung Neo TVs, and HomePod \cite{bthread}.

To ensure the security of a Thread network, only authorized devices can join the network, and all data communications inside the network are encrypted.
The security of the Thread network devices is ensured via two features, 
{\em viz.}\xspace, {\em authentication} and \textit{credential confinement}(\cite{bthreadspec}, p.1-4). 
Every new device is authenticated before being connected to the network; an encryption key is established to secure subsequent communication between the 
authenticated devices on the network.
The network-wide credentials (all the security and operational parameters used by the devices in the Thread network for dataset dissemination) must be confined to only the Thread devices. This credential confinement prevents any external device from receiving the network credentials, maintaining their confidentiality and ensuring the network's security.

Thread relies on its sub-protocols named \meshcopfull (MeshCoP) for secure authentication, commissioning, and letting join in new, untrusted devices to an existing mesh network.
MeshCoP uses the Datagram Transport Layer Security protocol (\dtls)\cite{bdtls} for message transferring.
It consists of three significant protocols, i.e., the \protocolcomm, which is responsible for authentication and registering of commissioner candidates; the \protocolthreadmanagement (\textsc{tmp}), which is for commissioner petitioning and management functionalities; and the \protocoljoiner, which is responsible for authentication, provisioning (a process of associating a device to its service) and entrusting (transferring network credentials) of the new untrusted Joiner device.

Given the widespread adoption of the Thread protocol, ensuring the security of its specification and implementation is crucial. Previous efforts have primarily focused on identifying bugs in Thread implementations through analyses of OpenThread \cite{bopenthred, bakestofidis}, examining side-channel vulnerabilities \cite{bdanialdinu}, and establishing security assessment taxonomies \cite{bliu2018}. However, to the best of our knowledge, no previous work has formally modeled the Thread protocol specification itself.

In this study, we address this gap by formally modeling MeshCoP, which represents a security-specific section of the Thread protocol. We then subject the model to a formal symbolic analysis against a set of security properties. To achieve this objective, we use ProVerif \cite{bproverif}, an automatic verification tool.


Our analysis targets version 1.1.1 of the Thread standard, followed by most IoT devices. 
We use its specification \cite{bthreadspec} as the basis of our modeling, in which cryptographic primitives are modeled as perfect black boxes, and messages are terms over these primitives.
The model is then analyzed against the Dolev-Yao attacker model \cite{bdymodel} to check security-relevant properties such as authentication, secrecy, Joiner trust, and key consistency.

\noindent\textbf{Contributions.} 
This work presents a case study of formal modeling and verification of MeshCoP using symbolic analysis.
We summarize our contributions as follows. 
\begin{itemize}
	\item \textbf{Modeling MeshCoP of Thread}. 
	We provide a formal model of MeshCoP, which can be used in related studies in the future. 
	Our model can serve as a component of a broader model of  Thread, facilitating Thread's formal modeling and analysis.
	In particular, we provide modeling abstractions for complex mathematical computations to generate and verify Schnorr’s zero-knowledge proofs in the authentication of the devices.
	Some of these modeling abstractions were essential to terminate proof queries in ProVerif.

	\item \textbf{Security analysis}. 
	We conduct an elaborate security analysis of the MeshCoP protocol, covering a wide range of its security-related properties, such as the secrecy of messages transferred, secure key derivation, authentication, and secure transfer of network credentials.
	
	Overall, our efforts enable checking 78 queries, including 40 for reachability, 15 for correspondence, and 23 for secrecy.
\end{itemize}

The paper is organized as follows. Section II covers MeshCoP background and symbolic modeling in ProVerif, and Section III covers modeling details, crypto primitive assumptions, modeling remarks, and handshake message transfer. 
Section IV presents the query results. Finally, 
Section V summarises Thread and ProVerif security analysis studies,
followed by conclusions in Section VI.
%
%
%
%
%
%
%
%
%
%
%
%
%
%
%

\section{Background}

This section provides a description of MeshCoP and symbolic modeling in ProVerif. 

\subsection{Protocol Description}

Let us first provide a simple description of MeshCoP.
MeshCoP consists of three sub-protocols: \protocolcomm, \protocolthreadmanagement, and \protocoljoiner.
\protocolcomm is responsible for the authentication and registration of a new device, termed as \textit{Commissioner Candidate} using \dtls \cite{bdtls}.
Upon successful authentication, the Commissioner Candidate registers its identity with the Border Agent by sending a \commpetreq message via the established \dtls Session.
\textsc{tmp} is responsible for the messages of petitioning, keep-alive, and management sent and received by the newly registered Commissioner.
Joiner Protocol is responsible for new Joiner authentication, subsequent provisioning, finalization, and then the secure transfer of Network Credentials (known as Joiner Entrust mechanism) to the Joiner.

MeshCoP participation devices include Border Agent, Commissioner, Commissioner Candidate, Joiner, Joiner Router, and Leader.
Border Agent is a device capable of relaying the MeshCoP messages between a Commissioner and a Thread Network. 
For \protocoljoiner, a Commissioner acts as an authentication server for the new joining Thread device.
A Commissioner Candidate is a device intending to be a Commissioner in the Thread Network.
Joiner is a device added by a human administrator to a Thread Network.
Joiner Router is a secured, one-hop away device from the Joiner.
Leader is a single distinguished central arbiter of the Thread Network.
On-Mesh Commissioner is a combined role of Commissioner, Border Agent, and optionally Joiner Router.
A Native Commissioner does not possess Network Credentials.

MeshCoP uses multiple credentials, including Commissioning Credential, Commissioning Key (\pskc), Joining Device Credential, and Network Credential.
A Commissioning Credential is a passphrase used to authenticate a Commissioning Candidate. 
Commissioning Key (\pskc) establishes a secure session between the Commissioner and the Border Agent. 
Joiner Device Credential is used to establish a secure connection between the Commissioner and the Joiner.
Network Credentials consist of all the security parameters (Network Master Key and Mesh Local Prefix) of the Thread Network.

\figurename \ref{fig_comm} and \figurename \ref{fig_join} show the messages flow for \protocolcomm, \protocolthreadmanagement and \protocoljoiner, respectively.
\dtlsshskeshd collectively represents \texttt{\dtlsch, \dtlsske} and \texttt{\dtlsshd} messages.
Similarly, \dtlsccf represents \dtlscke, \dtlsccs and \dtlsfin collectively; 
\dtlscf represents \dtlsccs and \dtlsfin message.

\subsection{Symbolic Modeling using ProVerif}
ProVerif is a symbolic protocol verifier extensively used for security protocol analysis. 
It takes input models in typed pi-calculus, translates them into the abstract Horn clause representation, and then applies resolution to these clauses.
The input pi-calculus models given to ProVerif include protocol specifications and properties to evaluate.
It then manipulates various cryptographic primitives specified by rewriting rules or specific equations.
It can validate protocols with an arbitrary number of parallel sessions or the size of messages.

ProVerif validates various security properties, including confidentiality, correspondence, as well as certain equivalence features.
It computes a set of \textit{terms} (messages) that the adversary knows to verify protocols using a symbolic model.
A message is classified as a secret if it does not belong to this set of \textit{terms}. 
This set can be possibly infinite for two reasons: the adversary can construct terms of unbounded length, and/or the protocol under investigation can be executed an infinite number of times.

\textbf{Limitations of ProVerif}: The tool may not always terminate and can lead to false attacks.
It does not support associative operations, e.g. \textsc{xor} (exclusive-or) operation.

\begin{figure}[ht!]
    \centering
    \begin{minipage}{\linewidth}
        \small 
        \begin{flalign*}
        M,N &::= &\text{terms} \\
            &\quad x, y, z &&\text{variable} \\
            &\quad a, b, c, k, s &&\text{name} \\
            &\quad f(M_1, \dots , M_n) &&\text{constructor application} \\
        D &::= &\text{expressions} \\
            &\quad M &&\text{term} \\
            &\quad h(D_1, \dots , D_n) &&\text{function application} \\
            &\quad \text{fail} &&\text{failure} \\
        P,Q &::= &\text{processes} \\
            &\quad 0 &&\text{nil} \\
            &\quad \text{out}(N,M); P &&\text{output} \\
            &\quad \text{in}(N, x : T); P &&\text{input} \\
            &\quad P | Q &&\text{parallel composition} \\
            &\quad !P &&\text{replication} \\
            &\quad \text{new } a : T; P &&\text{restriction} \\
            &\quad \text{let } x : T = D \text{ in } P \text{ else } Q &&\text{expression evaluation} \\
            &\quad \text{if } M \text{ then } P \text{ else } Q &&\text{conditional}
        \end{flalign*}
    \end{minipage}
    \caption{ProVerif: Syntax of terms, expressions, and processes.}
    \label{fig:representation}
\end{figure}

%
%
%
%
%
%
%
%
%
%
%

\section{Modeling MeshCoP}

We model MeshCoP to address three basic security objectives: Secrecy, Authentication, and Integrity of message \textit{msg}, passed between principals \textit{X} and \textit{Y} in presence of adversary \textit{Adv}.

\subsection{Adversary Model}
ProVerif, a Dolev-Yao model, can intercept all messages transferred in public channels, compute new messages, and re-send intercepted messages.
A Dolev-Yao adversary accesses \textit{free} public \textit{names} and \textit{terms}, provided to them through a public channel.

It considers strong (ideal) secrecy of encrypted data and generated hash.

Consequently, the integrity of the security primitives is maintained, contingent upon the confidentiality of their associated keys from potential adversaries.
We restricted the adversary from the generation of a new \funczkp. 

Brute force, side-channel, and guessing attacks are beyond the scope of our modeling. 

\subsection{Notations used in ProVerif Model}
For clear and concise representation, we present the notations used to represent participating entities, sent and computed new \textit{terms}, \textit{evaluations}, and \textit{restrictions}.
Our model represents broadly two parts: \textbf{Commissioner Protocol} and \textbf{Joiner Protocol}.

\textbf{Commissioner protocol}. Commissioner Candidate, Border Agent and Leader are represented as macros \process{ccli}, \process{bsrv} and \process{leader}, respectively.
Term \textit{ssnkeycommissioner} represents derived \dtls shared session key in processes \process{ccli} and \process{bsrv}; \texttt{sspcommissioner} represents pre-shared key phrase used to derive \pskc; 
Names \textit{cr} and \textit{sr} represents Client and server randomness;
Gx represents the Generator for public keys from their corresponding private keys;
\CLIDH, \SRIDH represents masked \process{ccli} and \process{bsrv} \ID;
\texttt{X$_i$, X$_{i+1}$} represents public key pair for which we model Schnorr \funczkp in \texttt{ClientHello, ServerHello} and \texttt{KeyExchange} messages;
\texttt{noncea}	represents the nonce generated by \process{bsrv};
\COMMSESSION represents the Commissioning session id sent by \process{leader};
\COMMSESSIONRCV represents the Commissioning session id received by \process{bsrv}.

\textbf{Joiner Protocol}.
Joiner, Commissioner, Border Agent, and Joiner Router are represented as process \process{joiner}, \process{csrv}, \process{brtrelay} and \process{jrtrelay}, respectively.
Restricted names \texttt{crj,srj} represents Client and server randomness;
\texttt{Gj$_x$} represents Generator for public keys from their corresponding private keys;
\texttt{Xj$_i$,Xj$_{i+1}$} public keys for which we model Schnorr \funczkp in \texttt{ClientHello, ServerHello} and \texttt{KeyExchange} messages; \texttt{derivedkeyjoiner} represents derived derived pre-Shared key as analogous to \pskc; \texttt{ntcreds} represents Network Credentials; \texttt{iid} represents individual identification id generated as hash of \process{joiner ip} address\texttt{joinerip};
\JNIDH represents masked \process{joiner} \ID;
Private free name \texttt{kek} represents Key Encryption Key- a one-time key used to transfer network credentials securely.
\aesprf represents Pseudo-Random Function.
New name \texttt{cook} means cookies transferred; 
the private free name \texttt{Granted} represents the acceptance of the Commissioning request after Commissioner Authentication, respectively.
Type \texttt{skey} \& \texttt{pkey} represents private key \& its public key.

\textbf{Communication Channels}:
We model dedicated communication channels for each process pair as \texttt{\channelch} (for \process{ccli} and \process{bsrv}); \texttt{\channelchbtoc} (for \process{csrv} and \process{brtrelay}); \texttt{\channelchbtoL} (for \process{bsrv} and \process{leader}); \channelchjtoc (for \process{joiner} and \process{jrtrelay}); and \texttt{\channelchrtob} (for \process{jrtrelay} and \process{brtrelay}).

 \begin{figure}[h!]\footnotesize
	 \centering
	 \begin{lstlisting}[language=ml, breaklines=true, mathescape]
fun element_to_key(element):key[data,typeConverter].
fun key_to_element(key):element[data,typeConverter].
fun ID_to_bitstring(ID):bitstring[data,typeConverter].
fun random_to_bitstring(random):bitstring[data,typeConverter].
	\end{lstlisting}
 \caption{Type Converters used in our model.}
 \label{fig:zkp}
 \end{figure}

Henceforth, we use the notations mentioned above in the paper to refer to them respectively.

\subsection{Modeling Notes}
Our model has the following security notions and invariants. 
MeshCoP uses Schnorr's Zero-Knowledge proofs to establish a DTLS-based secured session between an unauthenticated new joiner and the existing Thread network.

\begin{itemize}
	\item \textbf{Keys generation and sharing}: We generate secret keys, pre-shared keys with respective key-generators, as terms in the main process.
	Constructor \texttt{\PKDF(\aesprf,sspcommissioner,\OTHERVARS);} generates \pskc.
	We model a public key for a private key \texttt{x}, as \texttt{exp(G$_1$,x)}.
	For \dtls session keys, we derive public and shared secret key pair using the function \texttt{xx(skey,psk)} and ephemeral keys using function \texttt{exp(G$_x$,x)}.
	For sharing of the keys, Processes \process{ccli} and \process{bsrv} are initialized with commissioning credentials (a pre-shared key represented as \textit{sspcommissioner}).
	Processes \process{csrv} and \process{joiner} are initialized with a pre-shared key for the device (represented by \textit{sspjoiner}).
	Further, we initialize the processes with their public-private key pairs.

	\item \textbf{Zero-knowledge Proofs Generation}:
	We model non-interactive Schnorr's \zkp as \texttt{zk(v,x,s)}, where \texttt{v,x} and \texttt{s} represent an ephemeral private key, contextual private key, and signature, respectively.
	We model signature r, \texttt{r=v-x*h\%n} computation as constructor \texttt{r(skey,skey,bitstring)}.
	
	A \funczkp is sent as a pair of Public key to be verified and its \funczkp, and is received by a verifier as \texttt{(x:element,proof:bitstring)}.

	\item \textbf{Zero-knowledge Proofs Verification}:
	
	Schnorr's \funczkp requires complex operations, which are unavailable in ProVerif to verify a given \funczkp by the verifier.
	Hence, we model the \funczkp verification as a check between secrets \texttt{v} and \texttt{x} in the received \funczkp pair.
	
	To avoid non-termination, we use \texttt{equation forall x:skey,y:skey;exp(exp(G$_1$,x),y) = exp(exp(G$_1$,y),x)} to capture the relationship between constructors involving \texttt{G$_1$}.
	
	For ClientKeyExchange and ServerKeyExchange, we model the generators \texttt{G$_a$} and \texttt{G$_b$} where \texttt{G$_a$=X$_1$+X$_2$+X$_3$} and \texttt{G$_b$=X$_1$+X$_3$+X$_4$} as \texttt{G$_x$=gen\_h(X$_1$+X$_2$+X$_3$+X$_4$)}. Hence, for complete execution of all the mentioned events in our model, we require to capture the relationship among constructors using \texttt{G$_a$} and \texttt{G$_b$}. We model the relationship as \texttt{equation forall a:skey,b:skey,c:skey,d:skey,x:skey, y:skey; exp(exp(gen\_h(exp(G$_1$,a),exp(G$_1$,b),exp(G$_1$,c),exp(G$_1$,d)),x ),y)=exp(exp(gen\_h(exp(G$_1$,a),exp(G$_1$,b),exp(G$_1$,c),exp(G$_1$,d)),y),x)}.  
	
	\item \textbf{Cookie generation and verification}: 
	We model cookies using constructor \texttt{ssign(random\_to\_bitstring(cr),\textsc{id}\_to\_(srid))}, signed with masked server \texttt{id}.
	The client extracts \texttt{cr} i.e., client random value, to verify if the integrity of the cookie is preserved.
	
\end{itemize}

\subsection{Process events}
Our model for MeshCoP contains several special events:
\begin{itemize}
	\item \texttt{joinerrcvck(cookie)} represents the fact that \process{joiner} receives \texttt{cookie} from \process{csrv}; 
	event \texttt{cclircvck(cookie)} represents the fact that \process{ccli} receives \texttt{cookie} from \process{bsrv}; 
	event \texttt{bsrvsntck(cookie)} represents the fact that \process{bsrv} sends received \texttt{cookie} to \process{ccli};
	event \texttt{csrvsntck(cookie)} represents the fact that \process{csrv} sent \texttt{cookie} to \process{joiner}.
	
	\item \texttt{bsrvssk(ssnkeycommissioner)} represents the fact that \process{bsrv} has generated \dtls session key \texttt{ssnkeycommissioner} in secure \dtls connection.
	Similarly, event \texttt{cclissk(ssnkeycommissioner)} represents that \process{ccli} has generated \texttt{ssnkeycommissioner}.
	
	\item \texttt{bsrvfin(sspcommissioner,\pskc,cr,sr,G$_x$,\ID\\\_to\_bitstring(\CLIDH),\ID\_to\_bitstring(\SRIDH))} represents the fact that \dtls is completed on \process{bsrv} side with \texttt{sspcommissioner, \pskc,cr,sr,pkey} and respective principal \ID.
    Similarly, events \texttt{bsrvbeg, cclibeg, cclifin, csrvbeg, csrvfin, joinerbeg} and \texttt{joinerfin}  are defined for \process{bsrv}, \process{ccli}, \process{csrv} and Joiner processes, respectively.
	
	\item \texttt{eventdskjnr(derivedkeyjoiner)} represents the fact that \process{joiner} derived \dtls secret key \texttt{derivedkeyjoiner} using \texttt{cr, sr} and \texttt{sspcommissioner}.
	\item \texttt{eventdskcmm(derivedkeyjoiner)} represents the fact that \process{csrv} derived \dtls secret key \texttt{derivedkeyjoiner} using \texttt{cr, sr} and \texttt{sspcommissioner}.
	\item event \texttt{\eventrcvcommresponse(commpetres,Granted)} represents the fact that \process{ccli} receives \commpetres with response \texttt{Granted} for its \commpetreq.
	\item \texttt{rcvcommkarsp(commkares,Granted)} represents the fact \process{ccli} receives response \commkares with \texttt{Granted} for its \commkareq.
	\item \texttt{\eventsntcommresponse(commpetres,leadpetreq,leadpetres,Granted)} represents that Process \process{bsrv} sent \commpetres for \commpetreq of \process{ccli}.
	\item \texttt{sntcommkarsp(commkares,leadkareq,leadkares,Granted)} represents that \process{bsrv} sent \commkares for \textsc{comm}\_\textsc{ka}.req of \process{ccli}. 
	\item \texttt{\eventleaderrep(\SRID,leadpetreq1,leadpetres)} represents that \process{leader}responds with \texttt{leadpetres} for the corresponding \texttt{leadpetreq1} and $\SRID$.
	Similarly, \texttt{\eventleaderrepka} signals the same for \texttt{leadkareq}.
	\item \texttt{\joinergtsnetcreds(n)} represents the fact that \process{joiner}, after successful authentication, gets Network Credentials \texttt{n}, as Joiner Entrust.
	\item \texttt{\evjrtrsendsnetcreds(n)} represents the fact that \process{jrtrelay} sends Network Credentials \texttt{n}, to \process{joiner}.
\end{itemize}

\subsection{Handshake Messages Transfer Details}

\textbf{Commissioner protocol}. We list the major messages generated and transferred by the participating entities of Commissioner Protocol as in \figurename \ref{fig_comm}:

\begin{enumerate}
	\item Commissioner Protocol starts with \process{ccli}'s \texttt{ClientHello} message \textbf{M1} as \texttt{cr}, Schnorr \funczkp pairs i.e. public key \texttt{X$_1$} and \texttt{X$_2$} with their respective \funczkp proofs to convince the verifier about the possession of the claimed public keys, and sends it to \process{bsrv}.
	\process{bsrv} receives the \texttt{ClientHello} and verifies its \funczkp pairs. 
	If verified, 
	generates a \texttt{cookie} and sends it to \process{ccli} as \dtls \texttt{HelloVerifyRequest} as shown in \textbf{M2}.
	
	\item \process{ccli} extracts and verifies the \texttt{cookie} from received message and sends the \texttt{HelloVerifyRequest} to \process{bsrv} as shown in \textbf{M3}.
	\process{bsrv} verifies the authenticity of the \texttt{cookie}; generates and sends \texttt{ServerHello} similar to \texttt{ClientHello} with different parameters \texttt{sr}, Schnorr \funczkp pairs for public keys \texttt{X$_3$} and \texttt{X$_4$}, \texttt{KeyExchange} message with its \funczkp and \texttt{ServerHelloDone} to the \process{ccli} as shown in \textbf{M4}.

 \begin{figure}[ht]
    \centering
    \includegraphics[scale=.4]{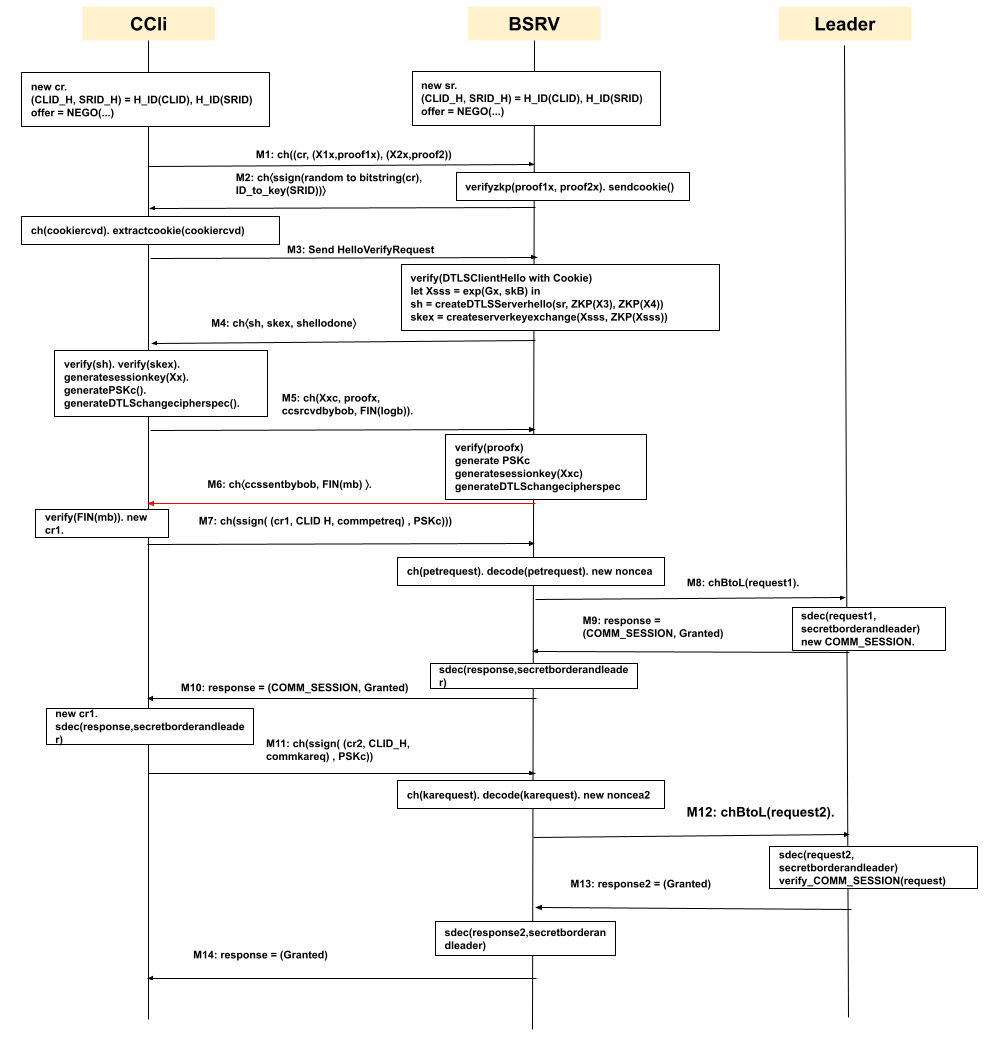}
    \caption{Overview of the Commissioner $\&$ Thread Management Protocol (\textsc{tmp}).}
    \label{fig_comm}
\end{figure}

\begin{figure}[ht]
    \centering
    \includegraphics[scale=.4]{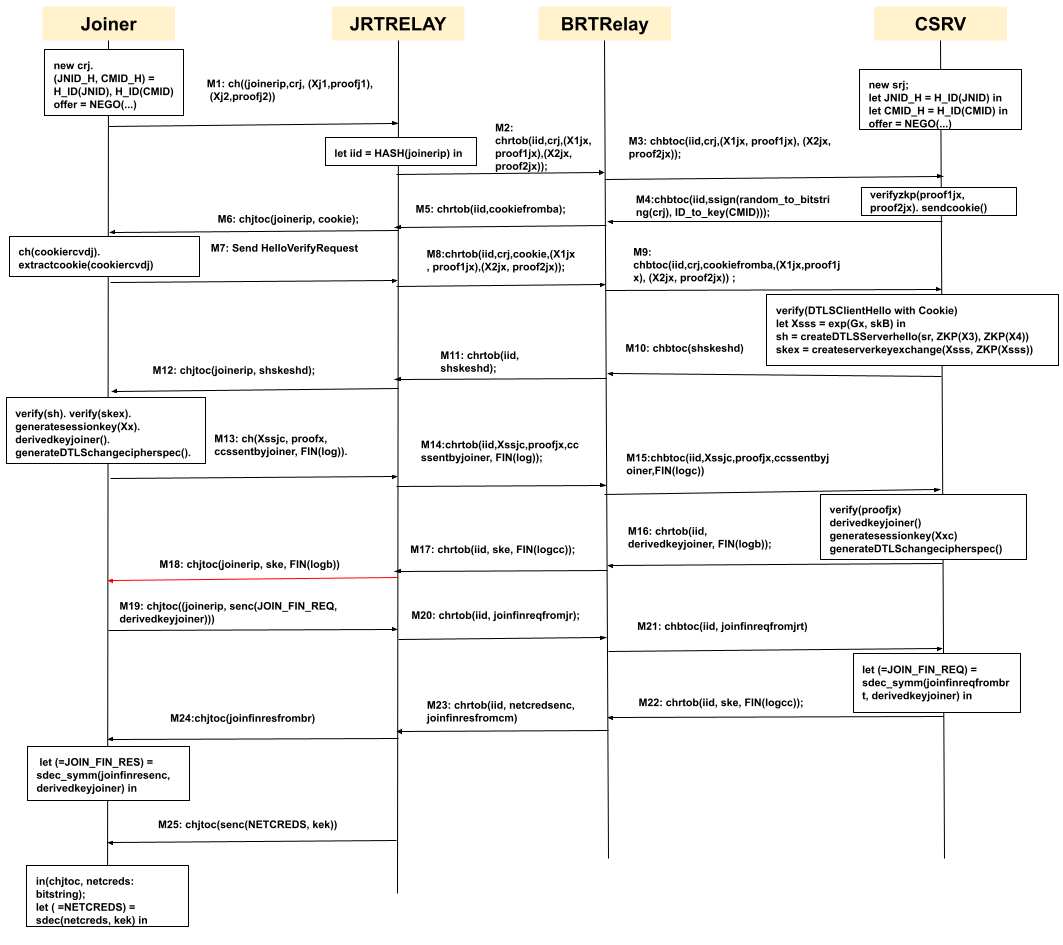}
    \caption{Workflow of the Commissioner $\&$ Thread Management Protocol (\textsc{tmp}). The red arrow shows the completion of Joiner Authentication.}
    \label{fig_join}
\end{figure}

	\item \process{ccli} verifies the received \funczkp pairs in \texttt{ServerHello} and \texttt{KeyExchange} message from \process{bsrv}. If successful, it derives \texttt{G$_x$}, generates Client \texttt{KeyExchange} message and \texttt{ChangeCipherSpec} as \texttt{ssign((cr,sr,\CLIDH),element\_to\_key\\(ssnkeycommissioner))}.
	Further, \process{ccli} derives and uses \pskc to generate \dtls \texttt{Finished} message;
	sends \texttt{KeyExchange}, \texttt{ChangeCipherSpec} and \dtls \texttt{Finished} to \process{bsrv} as shown in \textbf{M5}.
	\process{bsrv}, verifies Schnorr \funczkp of \texttt{Key Exchange} and generates \pskc \& \texttt{ssnkeycommissioner} to verify Client's \texttt{ChangeCipherSpec} message.
	Furthermore, it creates Server \texttt{ChangeCipherSpec} and \dtls \texttt{Finished} message similar to that by \process{ccli} and sends it to \process{ccli} as shown in \textbf{M6}.
	
	\item \process{ccli} receives and verifies the Server \texttt{ChangeCipherSpec} message; compares the contents of received \dtls \texttt{Finished} message with it's own prepared \dtls \texttt{Finished} message. 
	Hence, the Commissioner Authentication phase is completed.

	\item For Commissioner Registration phase, \process{ccli} signs \texttt{(cr1,\CLIDH,commpetreq)} with key \pskc, sends them to \process{bsrv} as shown in \textbf{M7}. 
	\process{bsrv} extracts the contents of the request.
	If successfully extracted, \process{bsrv} sends a \leadpetreq with generated name \texttt{noncea} as \texttt{(cr1,noncea,\textsc{clid}\_\textsc{h},leadpetreq)} as shown in \textbf{M8}.
	To keep the secrecy of messages transferred between process \process{bsrv} and \process{leader}, we encrypt them using a symmetric key \texttt{secretborderandleader}.

	Process \process{leader} receives encrypted \process{leader} petitioning request, decrypts it using \texttt{secretborderandleader}.
	The \process{leader} sends Leader petitioning response \texttt{leadpetres}, received nonce \texttt{noncea}, \texttt{srid}, a name \texttt{comm\_session} representing commissioning session id and a secret free name- \texttt{Granted}, representing Accept state) encrypted symmetrically using \textit{secretborderandleader} and then sends it to \process{bsrv} as in \textbf{M9}.
	\process{bsrv} extracts the received message contents and sends the \texttt{(cr,\CLIDH, commpetres,\textsc{comm\_sessionrcv},Granted)} to \process{ccli} as shown in \textbf{M10}.
	A similar approach is followed for \commkareq. 
	Hence, it finishes the Commissioner Protocol.
\end{enumerate}

\textbf{Joiner protocol}. Now, we explain the messages flow as given in \figurename \ref{fig_join}:
\begin{enumerate}
	
	\item Joiner protocol starts with \texttt{ClientHello} from message from \process{joiner} device containing names \texttt{joinerip,crj}, Schnorr \funczkp pairs i.e. public key \texttt{Xj$_1$} and \texttt{Xj$_2$} with their respective proofs, and sends it to \process{jrtrelay} as message \textbf{M1}.
	\item \process{jrtrelay} calculates \texttt{hash} of joiner ip address as variable \texttt{iid}, relays \texttt{iid} with \texttt{crj} and Schnorr \funczkp pairs to \process{brtrelay} as message \textbf{M2}, which is then forwarded to \process{csrv} as \textbf{M3}.
	\process{csrv} receives the \texttt{ClientHello} and verifies the \funczkp pairs; on successful verification, generates a \texttt{cookie} and sends it to \process{brtrelay} as \textbf{M4}, similar to Commissioner protocol, which relays it to \process{joiner} through \process{jrtrelay} as shown in \textbf{M5} and \textbf{M6}.
	
	\item On successful \texttt{cookie} verification, \process{joiner} relays the \texttt{HelloVerifyRequest} to \process{csrv}, through \process{jrtrelay} and \process{brtrelay} shown as \textbf{M7}, \textbf{M8} and \textbf{M9}.
	\process{csrv} verifies \texttt{cookie} and \texttt{ClientHello} message contents and if successful, generates \texttt{ServerHello} similar to ClientHello but with parameters \texttt{srj}, Schnorr \funczkp pairs for public keys \texttt{Xj$_3$} and \texttt{Xj$_4$}, \texttt{KeyExchange} message with it's \funczkp and \texttt{ServerHelloDone} to the joiner through \process{brtrelay} and \process{jrtrelay} as shown in \textbf{M10}, \textbf{M11} and \textbf{M12}.
	\process{joiner} verifies the received Schnorr \funczkp pairs in \texttt{ServerHello} and \texttt{KeyExchange} message.
	
	\item On successful verification, \process{joiner} derives \texttt{Gj$_x$} to generate Client \texttt{KeyExchange} Message, \texttt{ChangeCipherSpec} as \texttt{ssign((crj,srj,\JNIDH),element\_to\_key (ssnkeyjoiner))}.
	Further \process{joiner} derives \texttt{derivedkeyjoiner} \& generates \dtls \texttt{Finished} message as \texttt{ \textsc{fin}(ssign((\ID\_to\_bitstring($\JNIDH$),crj,srj, Gj$_x$),ssnkeyjoiner))}. Furthermore, \process{joiner} sends messages \texttt{KeyExchange}, \texttt{ChangeCipherSpec} and \texttt{DTLS Finished} to \process{csrv}, relayed through \process{brtrelay} and \process{jrtrelay} as shown in \textbf{M13}, \textbf{M14}, and \textbf{M15}.
	\process{csrv} verifies \funczkp of \texttt{Key Exchange} message and derives secret key \texttt{derivedkeyjoiner} as \texttt{get\_ms(crj, srj, sspjoiner)} and secret session key as \texttt{ssnkeyjoiner} to verify Client's \texttt{ChangeCipherSpec} message using the key \texttt{ssnkeyjoiner}.  
	Furthermore, it creates Server \texttt{ChangeCipherSpec} and \dtls \texttt{Finished} message similar to that by \process{joiner} and relays it to \process{joiner} as shown in \textbf{M16}, \textbf{M17} and \textbf{M18}.
	
	\item \process{joiner} receives and verifies the Server \texttt{ChangeCipherSpec} message; compares the contents of received \dtls \texttt{Finished} message with it's own \dtls \texttt{Finished} message. 
	Hence, \process{joiner} Authentication is completed.
	
	\item Upon successful Joiner Authentication, the \process{joiner} creates a Joiner Finalization request as a tuple \texttt{(joinerip,senc(\textsc{join}\_\textsc{fin}\_\textsc{req},derivedkeyjoiner))} and relays it request to \process{csrv} through \process{jrtrelay} and \process{brtrelay} as shown in \textbf{M19}, \textbf{M20} and \textbf{M21}.
	
	Receiver \process{csrv} decrypts the \joinfinreq with \texttt{derivedkeyjoiner}.\process{csrv} creates the \joinfinres as a tuple \texttt{(iid, \NETCREDS, kek)}, symmetrically encrypts it with key \texttt{scrtjtrcm} (to model secure message transfer inside the Thread network). \process{csrv} further sends \joinfinres to \process{brtrelay} which relays it to \process{jrtrelay} as shown in \textbf{M22} and \textbf{M23}. \process{jrtrelay} decrypts \joinfinres, extracts \texttt{kek} and then sends \texttt{\NETCREDS} (encrypted with \texttt{kek}) to \process{joiner} as \textbf{M24}. \process{joiner} receives and decrypts the message to extract \texttt{\NETCREDS} using \texttt{kek}.
	For simplicity, we combined \joinfinres with the Joiner Entrust signal.
\end{enumerate}
We have made our ProVerif model available online at \cite{modelcode} for reproducibility.

\section{Results}
Our ProVerif model consists of 40 reachability, 15 correspondence, and 23 secrecy queries which takes an average execution time of 13 minutes 48 seconds. The analysis is performed on a system with Intel(R) Xeon(R) W-1270 CPU @ 3.40GHz and 64 GB RAM.

\subsection{Correspondence Properties}
Thread specification describes the security properties informally.
We focus on parts that affect new Joiner candidates' authentication and secure credentials transfer.
In detail, we interpret the security requirements of MeshCoP in terms of the below-listed properties:

\begin{itemize}
	
	\item[Q1] \textbf{Cookie Consistency}: As the specification mentions \textit{"The Commissioner, upon reception of a \texttt{DTLS-ClientHello} message with a \texttt{cookie} from one of its scanned Joiners..."} [pp 8-29], it means \texttt{cookie} received by a client is sent by a genuine server. 
	\queryitem{Q1a} and \queryitem{Q1b} represents the Query for \protocolcomm and \protocoljoiner respectively to verify that property.
	\item[Q2] \textbf{Keys Consistency}: \queryitem{Q2a} verifies consistency of derived session keys and \pskc in \protocolcomm.
	\queryitem{Q2b} verifies the consistency of derived session keys and 
	\queryitem{Q2c} verifies consistency of derived secret keys, analogous to \pskc, in \protocoljoiner.
	\item[Q3] \textbf{Joiner Entrust}: According to the specification, \textit{"Once the Commissioner has authenticated a Joiner, it proceeds to signal to the Joiner Router that the Joiner is to be entrusted with the Network Credentials."} [pp 8-30]
	\item[Q4] \textbf{Authentication of untrusted client and trusted server in MeshCoP.} We model it as \textit{only after successful authentication of the new \process{joiner}, \process{csrv} signals the \process{jrtrelay} for Joiner Entrust.}
	\queryitem{Q4a} and \queryitem{Q4b} represents authentication of \process{ccli} and \process{bsrv} in Commissioner protocol.
	\queryitem{Q4c} and \queryitem{Q4d} represents authentication of \process{ccli} and \process{bsrv} in Commissioner protocol.
	\item[Q5] \textbf{Commissioner Registration}: As per specification, \textit{"\commpetres is sent as a separate response by \process{brtrelay} on receipt of \leadpetres from the \process{leader}"} as given in pp.8-45~\cite{bthreadspec}, which is verified in \queryitem{Q5a}.
	\queryitem{Q5b} verifies if \textit{"\commkares is sent as a separate response by \process{brtrelay} on receipt of \leadkares from the \process{leader}"} pp.8-46~\cite{bthreadspec}.
	\item[Q6] \textbf{Commissioner Petitioning}: \queryitem{Q6a} verifies that, \textit{"\leadpetres is sent from the Leader to the Border Agent in response to \leadpetreq Communicates whether or not the Leader will let the petitioning Commissioner Candidate take the role"} pp.8-69 ~\cite{bthreadspec}.
	\queryitem{Q6b} verifies \textit{"\leadkares is sent by the Leader to the Border Agent in response to a \leadkareq"} pp.8-69 ~\cite{bthreadspec}.
\end{itemize}

\subsection{Reachability and Consistency Queries}
In a ProVerif model, response to \texttt{query q:type; event(q)} as \texttt{false} makes sure that the execution trace exists at least at the point where \texttt{event q} is placed in the model. The expected execution result of such query is \texttt{false}.  We use the mentioned technique in reachability queries to verify the complete execution of our model. 
e.g. \texttt{query cook:bitstring; event(joinerrcvck(cook))}.

For consistency of derived terms, we query the conjunction of events with the transferred terms of the processes.
e.g., \texttt{query el:element;event(bsrvssk(el)) \&\& event(cclissk(el))}.
Similar events are used in other process macros to confirm the complete execution.
Furthermore, event queries are mandatory to get executed in the model since the correspondence properties depend on their successful execution.

In our model, all \texttt{event} queries are reachable for both Commissioner and Joiner protocol. 


\subsection{Secrecy Queries}
ProVerif solves a given secrecy query as a reachability problem i.e., \textit{to prove the secrecy of a term \textit{T}, check whether the attacker successfully reaches a state to derive a term T}.
The query should return \textit{false} if the queried term \textit{T} is secret; otherwise, \textit{true}.
We take two approaches for secrecy property verification:
\begin{enumerate}
	\item \textit{For data transferred between participants}:
	Free names e.g., \joinfinreq are sent in encrypted form in public channels.
	To verify the secrecy of such free names, we use \textit{query attacker}($\joinfinreq$).
	\item \textit{For derived data}:
	We model protocol entities to derive terms whose secrecy cannot be verified using the previous approach e.g. \pskc using function \textsc{pbkdf}.
	Hence, we define a free variable as \texttt{free secretpskc:bitstring[private]}, sent it encrypted to the adversary as \texttt{out(temp,senc(secretpskc,$\pskc$));}
	and verify the secrecy of \texttt{secretpskc} using the previous approach \texttt{1}.
\end{enumerate}

		

MeshCoP secrecy queries require the following \textit{free} or derived names to be kept secure:
For \protocolcomm,
free name \texttt{commpetreq, commpetres, commkareq} and \texttt{commkares},  representing CoAP Request \texttt{URI}, \commpetreq, \commpetres, \commkareq and \commkares, respectively, are expected to be confidential. 
 


For \protocolthreadmanagement,
\texttt{leadpetreq} (\leadpetreq) represents the request sent from the Border Agent to the \process{leader} for the Commissioner role; \texttt{leadpetres} (\leadpetres) indicates the acceptance of \process{ccl} as a Commissioner; \texttt{leadkareq} (\leadkareq) for the keep-alive request, sent periodically during the petitioning by \process{bsrv} to \process{leader}; \texttt{leadkares} (\leadkares) for the keep-alive response by \process{leader} to \process{bsrv} to keep the Commissioning session open.

For \protocoljoiner, 
\texttt{secretpskc, netcreds} representing \pskc and Network Credentials are expected to be kept secret. \joinfinreq, a request sent by \process{joiner} to get network credentials after successful authentication; \joinfinres, a response containing 'Accept' and \texttt{kek} attachment is expected to be kept secret.
Further, \texttt{kek}, a single-use pairwise key negotiated between the Joiner and the Commissioner, is expected to be kept secret. 
Furthermore, \texttt{sspjoiner\_sec} is used to check the secrecy of a pre-shared phrase (\texttt{sspjoiner}) to generate a key to transfer the data after Joiner Authentication, is required to remain secret.



\subsection{Correspondence queries}
A correspondence query captures the relationship as \textit{if an event e has been executed, then event e' has been executed before}. Consider, events \texttt{e(T$_1$,...,T$_n$)} and \texttt{e'(T$_1$,...,T$_n$)} has terms \texttt{M$_{1}$,...,M$_{j}$ and N$_1$,...,N$_k$} as arguments, are built by applying constructors to variables \texttt{t$_i$} of type \texttt{T$_i$}, in these events establish the relation between them. The resultant query is written as: \texttt{query t$_1$:T$_1$,...,t$_n$:T$_n$; event(e(M$_1$,...,M$_j$)) $\implies$ event(e'(N$_1$,...,N$_k$)).}

A stronger variant of correspondence query is in the form of the one-to-one relationship between events, known as \textit{injective-correspondence assertions}, replacing keyword \texttt{event} with \texttt{inj-event}; i.e., for each event \texttt{e}, there exists a distinct occurrence of event \texttt{e'} i.e. no single event \texttt{e}, can map to two or more events \texttt{e'}.
Table ~\ref{tab:correspondence} shows the results of correspondence queries.

\begin{table}[h!]
	\centering
	\caption{Correspondence Queries for MeshCoP: Expected and Actual result observed in model.
		Section 3.4 explains various parameters used in the query events.
	}
	\label{tab:correspondence}
	\begin{flushleft}
        \texttt{%
		\begin{tabularx}{\textwidth}{|l|X|l|l|}\hline
			\bfseries Query ID &  \bfseries Query & \bfseries Expected & \bfseries Actual  \\\hline
			Q1a &event(cclircvck(cook))$\implies$event(bsrvsntck(cook)). & true   & true\\\hline
			Q1b & event(joinerrcvck(cook))$\implies$event(csrvsntck(cook)). & true  & true  \\\hline
			Q2a & event(\eventBSrvsharedsecretkey(el,k))\&\& event(\eventCClisharedsecretkey(el,k)) & false & false \\\hline
			Q2b & event(csrvssk(e1))\&\&event(joinerssk(e1)) & false & false \\\hline
			Q2c & event(eventdskjnr(a)) \&\& event(eventdskcmm(a)) & false & false \\\hline
			Q3 & event(\joinergtsnetcreds(ntcreds))$\implies$ (event(\eventcsrvfin(p,c,s,eg,cl,sl)) \&\& event(\evjrtrsendsnetcreds(ntcreds)) \&\&  event(\eventjoinerfin(p,c,s,eg,cl,sl))) & true & true \\\hline
			Q4a & inj-event(\eventcclifin(ss,ds,cr,sr,eg,cl,sl))$\implies$inj-event(\eventbsrvbeg(ss,ds,cr,sr,eg,cl,sl)) & true & true \\\hline
			Q4b & inj-event(\eventbsrvfin(ss,ds,cr,sr,eg,cl,sl))$\implies$inj-event(\eventcclibeg(ss,ds,cr,sr,eg,cl,sl)) & true & true \\\hline
			Q4c & inj-event(joinerfin(p,cr,sr,eg,cl,sl))$\implies$inj-event(csrvbeg(p,cr,sr,eg,cl,sl)). & true & true \\\hline
			Q4d & inj-event(csrvfin(p,cr,sr,eg,cl,sl))$\implies$inj-event(joinerbeg(p,cr,sr,eg,cl,sl)) & true & true \\\hline
			Q5a & inj-event(rcvcommrsp(r3,req,mrq))$\implies$inj-event(sntcommrsp(nn,r3,req,lrq,lrs,mrq)). & true & true \\\hline
			Q5b & inj-event(rcvcommkarsp(r3,req,mrq))$\implies$inj-event(sntcommkarsp(nn,r3,req,lrq,lrs,mrq)). & true & true \\\hline
			Q6a & inj-event(sntcommrsp(nn,r3,req,lrq,lrs,mrq))$\implies$inj-event(leaderrep(r3,nn,id,lrq,lsrs)). & true & true \\\hline
			Q6b & inj-event(sntcommkarsp(nn,r3,req,lrq,lrs,mrq))$\implies$inj-event(leaderrepka(r3,nn,id,lrq,lrs)). & true & true \\\hline
		\end{tabularx}
    }
	\end{flushleft}
\end{table}

\subsection{Query Results}
Table \ref{tab:correspondence} shows the results of correspondence queries.
All the confidentiality measures have been proven to result in positive outcomes.
We mention the importance of \emph{equations} in the Zero-Knowledge proofs in Section 3.3, representing the algebraic relationship between the terms and helping the model to terminate. Further, we provide various constructors to represent the complex mathematical operations of the Schnorr \funczkp.
As it can be observed in \ref{tab:correspondence}, injective queries contain \textit{nonces}, which help to satisfy the given one-to-one correspondence in the events. An omission of these \textit{nonces} violates the queries, evidencing that the specification designers have missed explicitly specifying the nonces or left it as an implementation detail.

\textbf{Limitations of our work.} Symbolic model assumes perfect cryptography without considering the computational strength of primitives and adversaries
A broken primitive leads to a broken MeshCoP model. We assume the Thread Network to be secured and encrypted.

\section{Related Work}
\textbf{Security assessment of Thread.}
Akestoridis et al.\cite{bakestofidis} have performed a security analysis of Thread using the hardware and software tools used to analyze the Zigbee network.
They use the Thread-flashed development boards to successfully perform energy depletion and password-guessing attacks.
However, the experiment is confined to the scenario where the Commissioner and Joiner devices communicate directly without any relaying device.
Their work is a proof of concept implementation of the hypothesized attacks.
Dino et al.\cite{bdanialdinu} performed an electromagnetic side-channel vulnerability analysis of the Thread to get access to network credentials.
Liu et al.\cite{bliu2018} assess the security properties of Thread, applying the taxonomy of Building Automation Systems (\bas) protocols. 
Authors show the Thread, despite being more secure than the traditional non-IP based \bas protocols, has handshake messages jamming and key compromise vulnerabilities.
They provide a general overview of Thread security properties but lack formal implementation and supporting proofs.

\textbf{Applications of ProVerif for formal security analysis}.
Bhargavan et al. \cite{bbhargav2017} and \cite{bbhargav2022} have comprehensively analyzed \textsc{tls} 1.3 Draft-18 protocol to extend the guarantees of the \textsc{tls} to its implementations.
Authors use ProVerif to verify whether \textsc{tls} 1.3 prevents the well-known attacks of \textsc{tls} 1.2, if run in parallel with \textsc{tls} 1.2 and also study the effect of strong cryptographic primitives on the mechanical security verification of \textsc{tls} 1.3.
A study of confidentiality in protocols is performed in \cite{bArapinis} and \cite{bTedeschi}, while authentication properties of Simple Network Management Protocol (\textsc{snmpv3}) are studied in \cite{bAsadi}.
A formal analysis of OAuth 2.0 using ProVerif for login authorization, session validity, authorized data storage, and other properties is performed in \cite{bBansal}.
Analysis of routing protocols in \cite{bCortier2012} and \cite{bJohnson} using ProVerif; first automated security verification of properties- anonymity and unlinkability, for Yao’s two-party computation protocol, is verified in \cite{bBursuc} and \cite{bHirschi}.
Authors in \cite{bWang} have proved the security and authenticity properties of a proposed attribute-based credential system.
Formal verification of secrecy of applications data, forward secrecy of client and server data of transport protocol, and authenticity for protocols of Line encryption v1 using ProVerif is performed in \cite{bShi}.
Noise Protocol Framework security analysis is performed in \cite{bKobeissi}; verification of proposed BeleniosVS protocol with an unbounded number of genuine and forged voters voting for each candidate; privacy and receipt-freeness for multiple corruption scenarios is studied in \cite{bCortier2019}.

We identified the need for a formal security analysis of Thread.

\section{Conclusions and Future Directions}
In this study, we provide a symbolic model of MeshCoP, a sub-protocol of Thread that authenticates and commissions untrusted devices, using ProVerif. 
We demonstrate that secrecy and correspondence properties hold on 
the model.

We currently assume ideal security conditions for symbolic modeling and have abstracted away the computational aspects. Going forward, we
would enhance the current model with computations
and automating the synthesis of attacks in cases where security proofs fail.
Furthermore, we plan to extend this work to cover a larger subset of the Thread protocol, enabling a broader assessment of its security aspects.

\end{document}